\documentstyle[twocolumn,aps,floats,graphicx]{revtex}

\begin{document}

\newcommand{\bec}{\begin{center}}
\newcommand{\ec}{\end{center}}
\newcommand{\be}{\begin{equation}}
\newcommand{\ee}{\end{equation}}
\newcommand{\beqn}{\begin{eqnarray}}
\newcommand{\eeqn}{\end{eqnarray}}
\newcommand{\bet}{\begin{table}}
\newcommand{\ent}{\end{table}}
\newcommand{\bib}{\bibitem}

\wideabs{

\title{
Charge ordering and inter-layer coupling in cuprates 
}

\author{P. S\"ule} 
  \address{Research Institute for Technical Physics and Material Science,\\
Konkoly Thege u. 29-33, Budapest, Hungary,\\
sule@mfa.kfki.hu
}

\date{\today}

\begin{abstract}

\maketitle

 The inter-layer direct Coulomb coupling is analyzed in a charge ordered superlattice bilayer 
model in which pairing is supported by inter-layer Coulomb energy gain (potential energy driven 
superconductivity).
The 2D pair-condensate can be characterized by a charge ordered state
with a "checkerboard" like pattern seen by scanning tunneling microscopy.
The $2D \Leftrightarrow 3D$ quantum phase transition of the hole-content at $T_c$,
supported by $c$-axis optical measurements, is also studied.
The pair condensation might lead to the sharp decrease of the normal state $c$-axis anisotropy of the hole content 
and hence to the decrease of inter-layer dielectric screening.
The drop of the $c$-axis dielectric screening can be the primary source of the condensation energy below $T_c$.
We find that a net gain in the electrostatic energy occurs along the $c$-axis, which is proportional to the measured condensation energy ($U_0$) and with $T_c$:
$E_c^{3D} \approx 2 (\xi_{ab}/a_0+1)^2 U_0 \approx k_B T_c$ and is due to inter-layer charge complementarity (charge asymmetry of the boson condensate)
where $\xi_{ab}$ is the coherence length of the condensate and $a_0 \approx 3.9 \AA$ is the in-plane lattice constant.
The model naturally leads to the effective mass of $m^{*} \approx 4 m_e$ found by experiment.
The static $c$-axis dielectric constant $\epsilon_c$ is calculated for various cuprates
and compared with the available experimental data. 
We find correlation between $T_c$ and the inter-layer spacing $d$, $\epsilon_c$
and with the coherence area of the condensate.

\end{abstract}
}

%\vspace{2cm}

\section{Introduction}

 It is more or less generally accepted now that the conventional electron-phonon pairing mechanism 
cannot explain cuprate superconductivity, because as high a transition temperature as
$164 K$ (the record $T_c$ up to now \cite{Xiong}) cannot be explained by the energy scale of lattice vibrations without leading to lattice instability \cite{Hirsch}.
It is already well established that much of the physics related to high-temperature superconductivity (HTSC)
is in 2D nature,
one of the basic questions to be answered, however, in the future is whether HTSC is
a strictly 2D phenomenon or should also be described by a 3D theory.

 A well known experimental fact is that the zero resistance occurs along the $ab$ plane and the $c$-axis
at the same critical temperature \cite{Plakida} suggesting that there must be inter-layer (IL) coupling involved in the mechanism which drives the system into HTSC.
Another experiment, such as the intercalation on Bi2212 \cite{Choy}, however leads to the
opposite conclusion. Intercalation of I or organic molecules, which expands the unit cell
significantly along the $c$-axis, does not affect $T_c$. 
This finding is against IL coupling and supports low-dimensional theories.
The large anisotropy of the resistivity (and of other transport properties) is again not in favour of 3D theories of HTSC \cite{Plakida}. 
%Moreover, Basov {\em et al.} reported $c$-axis optical results and detected a small lowering of kinetic
%energy in the IL transport of cuprates \cite{Basov}.

  Experiments on $YBa_2Cu_3O_{7-y}$ (YBCO) ultrathin artificial HTSC compounds, sandwiched between thick nonsuperconducting
 $PrBa_2Cu_3O_{7-y}$ (PBCO) layers \cite{Plakida,Li} and measurements on $(BaCuO_{2+x})_2/(CaCuO_2)_n$
heterostructures \cite{Balestrino} indicate the continous decrease of $T_c$ with the decreasing number of
superconducting (SC) layers.
In particular the one-unit-cell thick sample of YBCO/PBCO heterostructure exhibits 
$T_c \approx 20$ K \cite{Li,Terashima}.
Other heterostructures, such as the $(Ba_{0.9}Nd_{0.1}CuO_{2+x})_5/(CaCuO_2)_2/
(Ba_{0.9}Nd_{0.1}CuO_{2+x})_5$, containing a {\em single bilayer SC block} isolated from each other by insulating blocks, were shown to have $T_c \approx 55$ K
\cite{Balestrino2}.
  These findings again indicate the importance of IL coupling in high temperature superconductors.
There are a couple of other findings against and supporting the 3D nature of HTSC \cite{Anderson2,PWA,Schneider}.
Most notably the systematic dependence of the transition temperature $T_c$ on the $c$-axis
structure and, in particular, on the number of $CuO_2$ planes in multilayer blocks are also
strongly in favour of the 3D character of HTSC. 
It is therefore, a fundamental question whether
at least a weak IL coupling is needed for driving the system to a superconducting (SC) state or a single $CuO_2$
layer is sufficient for HTSC \cite{Wheatley}.

There has been considerable effort spent on understanding HTSC within the context
of IL coupling mechanism in the last decades in which $c$-axis energy is available as a pairing mechanism
\cite{Mills,Tesanovic,Rajagopal,Xiang}. In other approaches the importance of IL hopping is
emphasized
{\em vs.} the direct IL Coulomb interaction of charged sheets \cite{Anderson2,Wheatley}.
There is another theory providing explanation for HTSC using the general framework of BCS combined
with IL coupling \cite{Ihm}.
The so-called IL tunnelling (ILT) theory \cite{Anderson2,Wheatley}, however, is no longer considered
a viable mechanism for SC in cuprates since ILT could provide no more then $1 \%$ of the
condensation energy in certain cuprates \cite{Hirsch,PWA,Tsvetkov}.

  In this paper we propose a simple phenomenological model for explaining the 3D character of HTSC
in cuprates supported by calculations.
We would like to study the magnitude of direct Coulomb interaction
between charge ordered square superlattice layers as a possible source of pairing interaction.
Our intention is to understand HTSC within the context of an IL Coulomb-mediated mechanism.
Of particular relevance to our investigation 
are the doping, multilayer, and pressure dependence of $T_c$ in terms of a charge ordered  {\em superlattice nature} of pair condensation.

  The size of a charge ordered characteristic superlattice can directly be related to the in-plane coherence length $\xi_{ab}$ of cuprates, which 
is proportional to the linear size of the pair condensate (real space pairing) in the {ab}-plane \cite{Tinkham}.
The relevant length scale for superconductors is the characteristic size of the Cooper-pair
and can be estimated by un uncertainty-principle argument \cite{Tinkham}.
Only those charge carriers play a decisive role in HTSC which has energy within 
$\sim k_B T_c$ of the Fermi energy and sets in at $T_c$
with the momentum range $\Delta p \approx k_B T_c / v_F$,
where $v_F$ is the Fermi velocity, leading to the definition of the
Pippard's characteristic length \cite{Tinkham}
\be
\xi_{ab} \approx a \frac{\hbar v_F}{k_B T_c}
\label{Pippard}
\ee
where $a$ is a numerical constant of order unity, to be determined.
$\xi_{ab}$ is a relevant number for describing the 2D confinement of the pair-condensate
wave function below $T_c$.
$\xi_{ab}$ can also directly be obtained according to the anisotropic Ginsburg-Landau theory
via the measurement of the upper critical field $H_{c2,ab}$ \cite{Tinkham}.
It has been shown recently that $\xi_{ab}$ obtained from Eq.~(\ref{Pippard})
or from $H_{c2,ab}$ measurements are very close to each other \cite{Wang_sci}.

 The IL charging energy we wish to calculate depends then on the IL spacing ($d$), the IL dielectric
constant $\epsilon_c$, the hole content $p$ and the size of the superlattice.
A large body of experimental data are collected which support our model.
A consistent picture is emerged on the basis of the
careful analysis of this data set. 
Finally we calculate the static $c$-axis dielectric constant $\epsilon_c$ for various cuprates
which are compared with the available experimental observations.

\section{Optimal doping and the 2D-3D phase transition of the hole-content}

 It is commonly accepted that charge carriers are mainly confined to the 2D $CuO_2$ layers
and their concentration is strongly influenced by the doping agent via hole
doping. Holes (no charge and spin at a lattice site) in the 2D $CuO_2$ layers are the key superconducting elements in high temperature superconductivity (HTSC). A characteristic feature of many high temperature superconductors (HTSCs) is the
optimal hole content value of $p_o \approx 0.16e$ per $CuO_2$ layer at optimal doping measured in the normal state (NS) \cite{Zhang}.
This general feature of cuprates can be summarized in the parabolic
 dependence of the maximum critical temperature $T_c^m$,
 \be
 T_c/T_c^m=1-82.6 (p-p_0)^2,
 \ee
where $T_c^m$ corresponds to the optimal hole concentration $p_0$ \cite{Tallon}.
Since $T_c$ appears to be maximized at $p_o \approx 0.16$, we pay special attention
to the IL Coulomb interaction in cuprates at this particular hole concentration.

{\em Our starting hypothesis is that
the hole content goes through a reversible $2D \Leftrightarrow 3D$ quantum phase transition 
at $T_c$ in layered cupper oxides.}
Optical studies on the $c$-axis charge dynamics reveals this phenomenon: the $c$-axis reflectance is nearly insulating
 in the NS but below $T_c$ is dominated by the Josephson-like plasma edge \cite{Basov}.
Below $T_c$ a sharp reflectivity edge is found at very low frequencies (lower then the superconducting gap) for a variety of cuprates
arising from the carriers condensed in the SC state to the $ab$-plane and due to the onset
of a coherent charge transport along the $c$-axis \cite{Tamasku,Tajima,Katz,Basov2,Motohashi}.
The appearance of plasma in the SC state and the absence of it in the NS seem to be
a common feature of HTSC cuprates \cite{Basov2}.
Band structure calculations predict an appreciable $c$-axis dispersion of bands close
to the Fermi surface and thus an anisotropic three-dimensional metallic state \cite{Pickett}.
Other first principles calculations indicate that
above $T_c$ the hole charge $p$ is charge transferred to the doping site in the charge reservoir ($2D \rightarrow 3D$ transition) \cite{Gupta,Sule}.

  Furthermore, we expect that
below $T_c$ the hole-content is condensed
to the sheets forming {\em anti-hole regions} (hole-content charge at a lattice site, $3D \rightarrow 2D$ transition). 
An anti-hole corresponds to an excess charge condensed to a hole lattice site in the
sheets below $T_c$.
Therefore, in the normal state hole doping is the
dominant charge transfer mechanism while in the superconducting (SC) state the opposite is true: the IL hole-charge is transferred back to the $CuO_2$ planes (charging of the sheets).
The latter mechanism can be seen by the measurement of transport properties as a function
of temperature \cite{Plakida,Ando,Semba,Yamamoto} if we assume that the increase in the density of
in-plane free carriers below $T_c$ is due to pair condensation.
The sharp temperature dependence of the $c$-axis dielectric constant $\epsilon_c$
and optical conductivity \cite{Kitano} 
seen in many cuprates and in other perovskite materials also raises the possibility of
a $2D \Leftrightarrow 3D$ charge density condensation mechanism at $T_c$ \cite{Plakida,diel_t}.
Therefore, the pair condensation can be described by an anisotropic 3D condensation mechanism,
and by a doping induced 2D-3D dimensional crossover \cite{Schneider}.

 In any naive model of electron pairing in cuprates the Coulomb repulsion is troublesome.
When the pairs of charged carriers are confined to the sheets, naturally a net self-repulsion
of the pair condensate occurs. Although short-range Coulomb screening in the dielectric crystal
can reduce the magnitude of the repulsion but it is insufficient to cancel
Coulomb repulsion \cite{Friedberg91}. 
Spatial separation should reduce the interaction strength, but even at $14 \AA$
 $e^2/r \approx 1 eV$ if unscreened.
One can assume capacitative effect between the $CuO_2$
planes: The charged boson condensate in one plane is stabilized by a deficiency of that charge
on another plane \cite{Friedberg91}. 
The various forms of the capacitor model are associated with the 3D character of HTSC, considering the inter-layer charge reservoir as a dielectric medium \cite{Friedberg91}.
 The inter-layer charging energy might be insufficient to stabilize the self-repulsion of the holes
in one plane and the self-repulsion of the charge condensate on another plane in the capacitor model.
Also, the superconducting properties of the hole-rich and nearly hole-free sheets would be different which is not verified by experimental techniques.

 Our intention is to combine the 2D and 3D nature of HTSC.
  Therefore, we propose a phenomenological model in which the charge distribution of the planes is polarized 
in such a way that holes and anti-holes (hole-electron pair) are 
phase separated within each of the sheets leading to a {\em charge ordered state} (COS).
Recently, signatures of charge ordering have been found in various cuprates and
manganites in the presence and the absence of magnetic fields \cite{Hoffman,Lake,Howald,Dagotto}.
Furthermore, the "checkerboard" charge pattern seen in Bi2212 \cite{Hoffman}
with spin periodicity of $\sim 8a_0$ strengthens our expectation that
the SC state can be characterized by a COS.
We focus on, then the magnitude of IL coupling (direct IL Coulomb interaction) between 2D static charge ordered superlattices.

\begin{figure}

\setlength{\unitlength}{0.07in}
\begin{picture}(20,20)(0,5)
\linethickness{0.35mm}
  \multiput(0,0)(5,0){5}{\line(0,1){20}}
  \multiput(0,0)(0,5){5}{\line(1,0){20}}
%  \multiput(0,0)(10,0){5}{\circle*{3}}
  \put(0,0){\circle*{5}}
  \put(0,5){\circle{3}}
  \put(0,10){\circle*{5}}
  \put(0,15){\circle{3}}
  \put(0,20){\circle*{5}}
  \put(5,0){\circle{3}}
  \put(10,0){\circle*{4}}
  \put(15,0){\circle{3}}
  \put(20,0){\circle*{4}}
  \put(5,5){\circle*{4}}
  \put(10,5){\circle{3}}
  \put(15,5){\circle*{4}}
  \put(20,5){\circle{3}}
  \put(5,10){\circle{3}}
  \put(10,10){\circle*{4}}
  \put(15,10){\circle{3}}
  \put(20,10){\circle*{4}}
  \put(5,15){\circle*{4}}
  \put(10,15){\circle{3}}
  \put(15,15){\circle*{4}}
  \put(20,15){\circle{3}}
  \put(5,20){\circle{3}}
  \put(10,20){\circle*{4}}
  \put(15,20){\circle{3}}
  \put(20,20){\circle*{4}}
%      \put(0,2.5){\circle*{2}}
%      \put(2.5,0){\circle*{2}}
%      \put(7.5,0){\circle{2}}
%      \put(5,2.5){\circle{2}}
%      \put(10,2.5){\circle*{1}}
%      \put(12.5,0){\circle*{1}}
\end{picture}

\setlength{\unitlength}{0.07in}
\begin{picture}(20,20)(-25,-15)
\linethickness{0.35mm}
  \multiput(0,0)(5,0){5}{\line(0,1){20}}
  \multiput(0,0)(0,5){5}{\line(1,0){20}}
%  \multiput(0,0)(10,0){5}{\circle*{3}}
  \put(0,0){\circle{3}}
  \put(0,5){\circle*{4}}
  \put(0,10){\circle{3}}
  \put(0,15){\circle*{4}}
  \put(0,20){\circle{3}}
  \put(5,0){\circle*{4}}
  \put(10,0){\circle{3}}
  \put(15,0){\circle*{4}}
  \put(20,0){\circle{3}}
  \put(5,5){\circle{3}}
  \put(10,5){\circle*{4}}
  \put(15,5){\circle{3}}
  \put(20,5){\circle*{4}}
  \put(5,10){\circle*{4}}
  \put(10,10){\circle{3}}
  \put(15,10){\circle*{4}}
  \put(20,10){\circle{3}}
  \put(5,15){\circle{3}}
  \put(10,15){\circle*{3}}
  \put(15,15){\circle{3}}
  \put(20,15){\circle*{4}}
  \put(5,20){\circle*{4}}
  \put(10,20){\circle{3}}
  \put(15,20){\circle*{4}}
  \put(20,20){\circle{3}}
%      \put(0,2.5){\circle*{2}}
%      \put(2.5,0){\circle*{2}}
%      \put(7.5,0){\circle{2}}
%      \put(5,2.5){\circle{2}}
%      \put(10,2.5){\circle*{1}}
%      \put(12.5,0){\circle*{1}}
\end{picture}

\setlength{\unitlength}{0.07in}
\begin{picture}(12,12)(0,-2.6)
\linethickness{0.35mm}
  \multiput(0,0)(5,0){5}{\line(0,1){20}}
  \multiput(0,0)(0,5){5}{\line(1,0){20}}
%  \multiput(0,0)(10,0){5}{\circle*{3}}
  \put(0,0){\circle*{4}}
  \put(0,5){\circle*{4}}
  \put(0,10){\circle*{4}}
  \put(0,15){\circle*{4}}
  \put(0,20){\circle*{4}}
  \put(5,0){\circle*{4}}
  \put(10,0){\circle*{4}}
  \put(15,0){\circle*{4}}
  \put(20,0){\circle*{4}}
  \put(5,5){\circle*{4}}
  \put(10,5){\circle*{4}}
  \put(15,5){\circle*{4}}
  \put(20,5){\circle*{4}}
  \put(5,10){\circle*{4}}
  \put(10,10){\circle*{4}}
  \put(15,10){\circle*{4}}
  \put(20,10){\circle*{4}}
  \put(5,15){\circle*{4}}
  \put(10,15){\circle*{4}}
  \put(15,15){\circle*{4}}
  \put(20,15){\circle*{4}}
  \put(5,20){\circle*{4}}
  \put(10,20){\circle*{4}}
  \put(15,20){\circle*{4}}
  \put(20,20){\circle*{4}}
%      \put(0,2.5){\circle*{2}}
%      \put(2.5,0){\circle*{2}}
%      \put(7.5,0){\circle{2}}
%      \put(5,2.5){\circle{2}}
%      \put(10,2.5){\circle*{1}}
%      \put(12.5,0){\circle*{1}}
\end{picture}

\setlength{\unitlength}{0.07in}
\begin{picture}(12,12)(-25,-15)
\linethickness{0.35mm}
  \multiput(0,0)(5,0){5}{\line(0,1){20}}
  \multiput(0,0)(0,5){5}{\line(1,0){20}}
%  \multiput(0,0)(10,0){5}{\circle*{3}}
  \put(0,0){\circle{3}}
  \put(0,5){\circle{3}}
  \put(0,10){\circle{3}}
  \put(0,15){\circle{3}}
  \put(0,20){\circle{3}}
  \put(5,0){\circle{3}}
  \put(10,0){\circle{3}}
  \put(15,0){\circle{3}}
  \put(20,0){\circle{3}}
  \put(5,5){\circle{3}}
  \put(10,5){\circle{3}}
  \put(15,5){\circle{3}}
  \put(20,5){\circle{3}}
  \put(5,10){\circle{3}}
  \put(10,10){\circle{3}}
  \put(15,10){\circle{3}}
  \put(20,10){\circle{3}}
  \put(5,15){\circle{3}}
  \put(10,15){\circle{3}}
  \put(15,15){\circle{3}}
  \put(20,15){\circle{3}}
  \put(5,20){\circle{3}}
  \put(10,20){\circle{3}}
  \put(15,20){\circle{3}}
  \put(20,20){\circle{3}}
%      \put(0,2.5){\circle*{2}}
%      \put(2.5,0){\circle*{2}}
%      \put(7.5,0){\circle{2}}
%      \put(5,2.5){\circle{2}}
%      \put(10,2.5){\circle*{1}}
%      \put(12.5,0){\circle*{1}}
\end{picture}

\vspace{-1.5cm}
\caption{\small Upper panels: The charge odered state of the type of a "checkerboard" of the hole-anti-hole condensate.
Opened and filled circles denote the holes with a charge of $q_i=+0.16e$ and anti-holes ($q_j=-0.16e$),
respectively in the $4a_0 \times 4a_0$ ($5 \times 5$) square lattice layer model.
%Each circle represent a $CuO_2$ unit cell. The anti-hole charges sum up to the
%charge of a boson pair $2e$ on this characteristic superlattice.
Note that the left panel accomodates $13$ anti-holes which corresponds to
$2e+0.16e/2$ charge. The right panel contains $12$ anti-holes corresponding
to $2e-0.16e/2$ charge.
The two lower panels correspond to the antiferromagnetic insulating state (left)
and to a non charge ordered holed-doped system (normal state, right).
}
\label{fig1}
\end{figure}
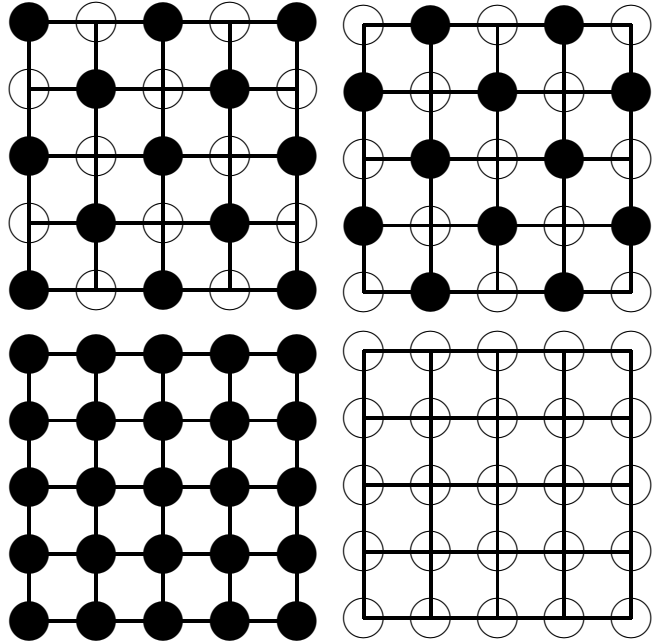

   {\em Charging of the sheets}:
 The $CuO_2$ layers carry negative charge obtained from the charge reservoir
even in the insulating stochiometric materials, e.g. in the infinite layer compound
$CaCuO_2$ each layer is charged by $2e$ charge donated by the $Ca$ atoms.
In $(CuO_2)^{2-}$ then the electron configuration of $Cu$ and $O$ is $3d^{10} 4s^1$ and 
$2p_z^2$. 
Unit $(CuO_2)^{2-}$ is typical of any undoped (stochiometric) cuprates and {\em ab initio} calculations
provide approximately the charge state $Cu^{1.5+}O_2^{3.5-}$ \cite{Plakida,Pickett}
which is due to charge redistribution between Cu and O.
The $(CuO_2)^{2-}$ plane itself is antiferromagnetic and insulating, e.g. there are no holes
in the "overcharged" layers.
Upon e.g. oxygen doping, however, the doping charge $p$ is transferred along the $c$-axis
to the doping site in the charge reservoir, since the doping atom exhibits a rare gas electron 
structure ($O^{2-}$, if oxygen doping) \cite{Gupta,Sule}.

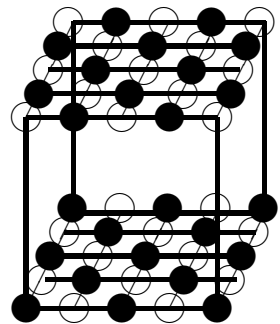
\begin{figure}

\setlength{\unitlength}{0.05in}
\begin{picture}(30,30)(-18,4)
%\thicklines
\linethickness{0.55mm}
  \put(0,0){\line(1,2){5.0}}
  \put(20,0){\line(1,2){5.0}}
  \put(0,20){\line(1,2){5.0}}
  \put(20,20){\line(1,2){5.0}}
  \put(5,0){\line(1,2){5.0}}
  \put(10,0){\line(1,2){5.0}}
  \put(15,0){\line(1,2){5.0}}
  \put(5,20){\line(1,2){5.0}}
  \put(10,20){\line(1,2){5.0}}
  \put(15,20){\line(1,2){5.0}}
  \put(20,0){\line(0,1){20.0}}
  \put(0,0){\line(0,1){20.0}}
  \put(0,20){\line(1,0){20.0}}
  \put(0,0){\line(1,0){20.0}}
  \put(5,30){\line(1,0){20.0}}
  \put(5,10){\line(0,1){20.0}}
  \put(25,10){\line(0,1){20.0}}
  \put(5,10){\line(1,0){20.0}}
  \put(2,3){\line(1,0){20.0}}
  \put(2.5,5.5){\line(1,0){20.0}}
  \put(4,8.){\line(1,0){20.0}}
  \put(1.3,22.5){\line(1,0){20.0}}
  \put(2.3,25.0){\line(1,0){20.0}}
  \put(3.5,27.5){\line(1,0){20.0}}
%  \multiput(0,0)(0,5){5}{\line(1,0){20}}
%  \multiput(0,0)(10,0){5}{\circle*{3}}
  \put(0,0){\circle*{3}}
  \put(5,0){\circle{3}}
  \put(10,0){\circle*{3}}
  \put(15,0){\circle{3}}
  \put(20,0){\circle*{3}}
%  \put(15,15){\circle*{8}}
  \put(0,20){\circle{3}}
  \put(5,20){\circle*{3}}
  \put(10,20){\circle{3}}
  \put(15,20){\circle*{3}}
  \put(20,20){\circle{3}}
  \put(1.6,3){\circle{3}}
  \put(6.3,3){\circle*{3}}
  \put(11.5,3){\circle{3}}
  \put(16.5,3){\circle*{3}}
  \put(21.3,3){\circle{3}}
  \put(2.5,5.5){\circle*{3}}
  \put(7.5,5.5){\circle{3}}
  \put(12.5,5.5){\circle*{3}}
  \put(17.5,5.5){\circle{3}}
  \put(22.5,5.5){\circle*{3}}
  \put(4,8){\circle{3}}
  \put(8.6,8){\circle*{3}}
  \put(13.8,8){\circle{3}}
  \put(18.8,8){\circle*{3}}
  \put(23.8,8){\circle{3}}
  \put(4.8,10.5){\circle*{3}}
  \put(9.8,10.5){\circle{3}}
  \put(14.8,10.5){\circle*{3}}
  \put(19.8,10.5){\circle{3}}
  \put(24.8,10.5){\circle*{3}}
  \put(1.3,22.5){\circle*{3}}
  \put(6.3,22.5){\circle{3}}
  \put(10.8,22.5){\circle*{3}}
  \put(16.4,22.5){\circle{3}}
  \put(21.4,22.5){\circle*{3}}
  \put(2.3,25.0){\circle{3}}
  \put(7.3,25.0){\circle*{3}}
  \put(12.3,25.0){\circle{3}}
  \put(17.3,25.0){\circle*{3}}
  \put(22.2,25){\circle{3}}
  \put(3.3,27.5){\circle*{3}}
  \put(8.5,27.5){\circle{3}}
  \put(13.2,27.5){\circle*{3}}
  \put(18.3,27.5){\circle{3}}
  \put(23,27.5){\circle*{3}}
  \put(4.4,30.0){\circle{3}}
  \put(9.4,30.0){\circle*{3}}
  \put(14.4,30.0){\circle{3}}
  \put(19.4,30.0){\circle*{3}}
  \put(24.4,30.0){\circle{3}}
\end{picture}

\vspace{1cm}
\caption{\small The charge ordered state of the hole-anti-hole condensate
in the $4a_0 \times 4a_0$ charge ordered bilayer superlattice model. Note the charge asymmetry between the adjacent layers. The bilayer can accomodate a pair of boson condensate ($4e$).
Noteworthy that holes (empty circles) and anti-holes (filled circles) can be characterized
by stripes along the diagonal lines.
The inter-layer charge complementarity of these charge stripes is crucial for the inter-layer Coulomb
energy gain.
}

\end{figure}

Briefly, the $c$-axis anisotropy of the hole content is strongly temperature dependent in cuprates.
  Our basic assumption is that in the SC state every second hole is filled up with the charge of $2p$
due to the {\em condensation} of the hole-content to the sheets leading to a charge ordered state in
the SC state.
Therefore the annihilation of the half of the holes results in the appearance of a  hole-anti-hole pair.
The hole-anti-hole pairs are the basic quasiparticles of the SC COS.
{\em The phase separation of the holes and anti-holes (leading to a charge ordered state) is stabilized by the intra- and inter-layer hole-anti-hole interactions.}
Basically the phase separation of holes and anti-holes in the planes is the manifestation of strong Coulomb
correlation in the SC state.
The superconducting charge-ordered states in cuprates are characterized theoretically based on the
assumption that cuprates are in close proximity to quantum critical points where
spin/or static charge order occurs in the SC state \cite{Vojta2}.
 In this paper we will focus on
the Coulomb interaction between the charge ordered sheets screened by the
IL dielectric media (charge reservoir).
It will be immediately apparent from our analysis that net energy gain occurs
due to the IL Coulomb interaction of the charge ordered layers when the hole-electron charge 
pattern is asymmetrically distributed in the adjacent layers (Fig. 2).

\section{The superlattice model: the coherence area}
  
 We propose to examine the following superlattice model of pair condensation:
 A pair of charge carriers ($2e$) van be distributed over $2/0.16=12.5$ $CuO_2$ unit cells in a square
lattice layer if the $2e$ pair is composed of the hole content $p_0 \approx 0.16/CuO_2$ at optimal doping.
However, allowing the phase separation of hole-anti-hole pairs, every second unit cell is occupied by
$-0.16e$ (anti-hole), and the rest is empty (holes, $+0.16e$), therefore we have $25$ unit cells for a condensed pair of charge 
carriers
 (Fig 1., that is the unit lattice of the pair condensate in the $5 \times 5$ supercell model).  
Therefore, $2 p_0 =-0.32e$ hole charge condenses to every second hole forming anti-hole sites.
The $25$ lattice sites provide 
the hole content of $25 \times 0.16e=4e$ and therefore $2e$ excess charge in the anti-hole sites.
In other words a Cooper wave-function can be distributed on a $5 \times 5$ superlattice
with a node of $-0.16e$ in every anti-hole site.
The remarkable feature is that the size of the $5 \times 5$ condensate (four lattice spacings, $4 a_0 \approx 15.5 \AA$) is comparable with the measured small coherence
length $\xi_{ab}$ of single-layer cuprates ($\xi_{ab} \sim 10 \AA$ to $20 \AA$) \cite{Plakida,Tinkham,Dagotto,Stinzingen}.
$\xi_{ab}$ can directly be related to the characteristic size of the wave-pocket of the local Cooper pair (coherence area) \cite{Plakida,Tinkham}.
The charge ordered superlattice model can in principle be applied not only for $p_0 \approx 0.16e$ but also for the entire doping regime.
The only difference
what happens is that the charge nodes in the charge ordered state changes
upon doping and consequently the inter-layer coupling scales with the anti-hole
charges in the underdoped regime. In the overdoped regime, however, the situation is more
difficult \cite{Tallon02,TallonLoram}. 

  Our expectation is that the COS of the $5 \times 5$ model given in Fig. 1 can be an effective 
model state for describing the SC state.
An important feature of this model is the {\em charge separation $dq$} in the charge ordered state,
where $+0.16e$ and $-0.16e$ partial boson charges are localized alternatively ($dq=0.32e$).
The hopping of charge carriers from the anti-hole sites to the holes can reduce the magnitude of the charge separation $dq$ leading to the extreme case when
$dq \approx 0$ which is nothing else then the $(CuO_2)^{2-}$ antiferromagnetic insulating state. 
Therefore characteristic quantity of the SC state $dq$ is directly related to the hole-content seen in cuprates in the
NS as a function of doping. 

 The hopping of anti-holes in such a charge ordered lattice layer might lead to SC without
considering any lattice vibration effects when $dq$ is in the optimal regime.
The freezing of the COS leads to insulating Wigner crystal (the NS) \cite{Vojta}. Therefore the melting of the crystallized COS is required for SC \cite{Vojta2}.
HTSC can be characterized in this way as a competition between a liquid-crystal-like striped COS (SC) and
a Wigner-crystal-like phase (NS) in a semiclassical theory of hole dynamics \cite{Vojta}.

 Below $T_c$ the charged-ordered state becomes stable compared with the
competing phase of the NS supported by IL charging energy.
Holes and anti-holes are placed in such a way in adjacent layers to maximize IL charging energy.
Therefore, holes in one of the layers are always covered by anti-holes in the other layer and {\em vice versa}
(FIG. 2, inter-layer {\em electrostatic complementarity}, bilayer $5 \times 5$ ($4a_0 \times 4a_0$) model).
An important feature is then that
the boson condensate can be described by an inter-layer {\em charge asymmetry}.
The IL coupling of the boson-boson pairs in the bilayer $5 \times 5$ model naturally suggests the
effective mass of charge carriers $m^{*} \approx 4 m_e$, as it was found by measurements \cite{Stasio,krusin}.
In the next section we generalize the $5 \times 5$ model to represent a real space periodicity of $N \times N$ coherence
area.
The charge ordered state presented in this article can also be studied as a charge density
wave (CDW) condensed on a superlattice.
Using the CDW terminology one has approximatelly two CDWs within a characteristic bilayer
with an amplitude of $\sim 0.16e$.
This system might not be insulating if we assume high rate of intersite hopping
and low rate of IL hopping of the charge carriers.
Then the charge ordered state is stabilized due to the attractive IL 
Coulomb interactions of asymmetrically condensed neighbouring CDWs.

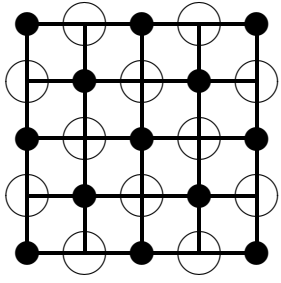
\begin{figure}

\setlength{\unitlength}{0.06in}
\begin{picture}(20,20)(-17,2)
\linethickness{0.35mm}
  \multiput(0,0)(5,0){5}{\line(0,1){20}}
  \multiput(0,0)(0,5){5}{\line(1,0){20}}
%  \multiput(0,0)(10,0){5}{\circle*{3}}
  \put(0,0){\circle*{2}}
  \put(0,5){\circle{4}}
  \put(0,10){\circle*{2}}
  \put(0,15){\circle{4}}
  \put(0,20){\circle*{2}}
  \put(5,0){\circle{4}}
  \put(10,0){\circle*{2}}
  \put(15,0){\circle{4}}
  \put(20,0){\circle*{2}}
  \put(5,5){\circle*{2}}
  \put(10,5){\circle{4}}
  \put(15,5){\circle*{2}}
  \put(20,5){\circle{4}}
  \put(5,10){\circle{4}}
  \put(10,10){\circle*{2}}
  \put(15,10){\circle{4}}
  \put(20,10){\circle*{2}}
  \put(5,15){\circle*{2}}
  \put(10,15){\circle{4}}
  \put(15,15){\circle*{2}}
  \put(20,15){\circle{4}}
  \put(5,20){\circle{4}}
  \put(10,20){\circle*{2}}
  \put(15,20){\circle{4}}
  \put(20,20){\circle*{2}}
%      \put(0,2.5){\circle*{2}}
%      \put(2.5,0){\circle*{2}}
%      \put(7.5,0){\circle{2}}
%      \put(5,2.5){\circle{2}}
%      \put(10,2.5){\circle*{1}}
%      \put(12.5,0){\circle*{1}}
\end{picture}

\vspace{1cm}
\caption{\small A possible charge ordered state of the normal state.
Opened and filled circles denote the holes with a charge of $q_i \approx +0.32e$ and anti-holes ($q_j \approx -0.16e$),
respectively in the $5 \times 5$ square lattice layer model.
}
\label{fig_ns}
\end{figure}

 We characterize the purely {\em hole-doped NS} as follows: In each sheet each $CuO_2$ units posses $p=+0.16e$ charge.
 The hole charge $p$ is charge transfered to the charge reservoir (doping site) above $T_c$.
 In this model one can naively expect that the self-repulsion of a hole rich sheet can be quite large. In order to reduce the hole-hole self-repulsion of the sheets, a hopping of $\sim p$ charge between the adjacent hole sites would result in the reduction of the repulsion leading to a Wigner-crystal-like COS depicted on FIG~\ref{fig_ns}. As is well known, the hopping of certain amount of charge ($\sim p$) between the holes leads to metallic conductance which is typical of hole-doped cuprates.
%Strong Coulomb forces also arise between the negatively charged doping site
%and the positive sheets. This interaction may well stabilize the system.
This is the {\em normal state} (NS) of the cuprates. 
Due to the strong $c$-axis anisotropy of the hole content in the NS 
IL dielectric screening is large in this state because of the increase in the dynamic component
of the IL dielectric permittivity ($\epsilon_c$) and therefore the IL interaction energy vanishes,
$E_c^{NS} \approx 0$.
That is the enhanced dielectric screening in the NS due to the localization of the hole content
in the charge reservoir. The lack of the IL plasma edge in the $c$-axis optical spectra
of various cuprates in the NS is a strong evidence of this phenomenon \cite{Molegraaf}.

\vspace{3cm}

\vspace{-3cm}

Our proposal is to understand
HTSC in such a way that the pair-condensate is stabilized at higher
temperature by the IL-charging energy provided by the antisymmetry of the
pair-condensate charge density between the adjacent layers.
 The $c$-axis dielectric constant of the SC state is reduced to the average value of
the IL dielectric medium due to the pair condensation of the hole charge to the sheets.
No IL screening of the hole charge occurs in this state which may well lead to energy gain
.

\section{The total energy of the pair condensate}

 Our intention is to develop a simple working hypothesis in which the IL charge reservoir
provides an average dielectric background and the hole content further enhances
IL dielectric screening when $2D \rightarrow 3D$ transition occurs (hole-doping, the NS).
The dielectric plasma provided by the hole charge results in the dynamic screening
effect of Coulomb interaction which is typical of hole doped cuprates in the NS.
In the opposite case ($3D \rightarrow 2D$, pair condensation) the IL dielectric screening
is nearly reduced to the average background value of the IL ion core spacer. This is what leads to IL energy gain.
The underlying source of the condensation energy is then the energy gain due to the lack of
dynamic screening in the SC state.
The possibility of direct IL hopping of Cooper pairs is not considered within this model as it was ruled out as
important aspect of HTSC \cite{PWA}.

We start from a very general description of our model system using e.g. a
Hamiltonian similar to that is given elsewhere \cite{Tesanovic} or which
can also partly be seen in general text books \cite{Kittel}.
We would like to describe then the
$2D \Leftrightarrow 3D$ condensation of the hole-content 
 using the Hamiltonian
\be
H=\sum_i H_i^{2D}+\sum_{i,j} H_{ij}^{3D},
\label{hamilt}
\ee
where $H_i^{2D}$ is the BCS-type Hamiltonian of the intra-layer condensate.
\beqn
H_i^{2D}=\sum_{k,\sigma} \epsilon_k c_{k \sigma,i}^{\dagger} c_{k \sigma,i}
\\ \nonumber 
+V \sum_{k,k'} c_{k\uparrow,i}^{\dagger} c_{-k\downarrow,i}^{\dagger} c_{k\downarrow,i}
c_{-k\uparrow,i}.
\label{intra}
\eeqn
In Eq. (4) $c_{k\uparrow,i}^{\dagger}$ is the creation operator for electrons in the
$i$th layer, with linear momentum $\bf k$ within the layer and spin $\sigma$,
using the effective-mass approximation $\epsilon_k=\hbar^2 k ^2/2 m^*$. 
$V= -\vert V \vert$ is the attractive inter-layer interaction (3D coupling) and is assumed
to originate from some of the proposed mechanisms \cite{Anderson2,Mills,Tesanovic,Rajagopal}.
The attractive IL term $H_{ij}^{3D}$ given by
\beqn
H_{ij}^{3D}=-t \sum_{k \sigma,\alpha,\beta} c_{k\sigma,i}^{\dagger} c_{k\sigma,j}
+ Y\sum_{k,k'} c_{k \alpha,i}^{\dagger} c_{-k \beta,j}^{\dagger} 
c_{-k \beta,j} c_{k \alpha,i} \\ \nonumber
+W\sum_{k,k',\alpha,\beta} c_{k \alpha,i}^{\dagger} c_{-k \beta,i}^{\dagger}
c_{-k \beta,j} c_{k \alpha,j}.
\eeqn
The first term in Eq. (5) is the direct IL hopping, while the second and third
terms describe IL coupling in a general form.
$t$ is very small in oxide superconductors \cite{PWA,Tesanovic}. $Y$ denotes
the IL coupling assisted by 
direct Coulomb interaction between charged layers (this is of particular
interest in our model).
$W$ is the coupling constant and can arise through the Coulomb interaction
causing inter-band transitions at the Fermi surface to some of the occupied or empty bands
away from the Fermi level, with finite dispersion, however, along the
$c$-axis.
In this paper we consider only the second term in Eq. (5) as the source of
the attractive interaction for pairing and neglect the rest of the inter-layer Hamiltonian
$H_{ij}^{3D}$ ($t \approx 0, W \approx 0$).

  We calculate then the energy of the nearly 2D electron pair condensate in the $CuO_2$ plane for the
bilayer supercell problem. 
It is assumed that the pair condensate behaves as a nearly free electron
gas with $ab$-kinetic energy $T_{kin}^{ab}$ and potential energy, which is mainly its in-plane self-electrostatic
energy ($E_c^{ab}$) and the out-of-plane inter-layer interaction energy $E_c^{IL}$.
For simplicity, the rest of the electron system is neglected completely.
The external potential of the condensate is also excluded in this model, that is the
lattice-condensate interaction, which is assumed to be negligible in the charged
$(CuO_2)^{2-}$ system (at least its contribution is negligible to the condensation energy).
In other words the ionic background of the planes is screened by the
core and valence electrons of the $(CuO_2)^{2-}$ plane. 
The kinetic energy of the charge condensate is due to the hopping of the charge carriers
between the $CuO_2$ sites within the sheets.
We do not take into account the complications due to ionic heterogeneity, nonpointlike polarization, etc.
The effective Coulomb interaction between spinless point-charges 
may be approximated by the expression $V_{eff}({\bf {r}})=e^2/(4 \pi \epsilon_0 \epsilon_c {\bf {r}})$, where $\epsilon_c$ takes into account phenomenologically the dielectric screening
effect of the IL dielectric medium and confined hole charge.
This kind of a rough approximation has widely used by several groups in the last decade
\cite{Mills,Friedberg91,Stasio,Ariosa}.
%and quantum and thermal effects. 
The important feature is that the $\underline{c}$-axis dielectric screening ($\epsilon_c$) is nearly reduced
to the static value of the average background dielectric constant in the SC state.
In accordance with this
a sharp temperature dependence of $\epsilon_c(\omega)$ is found in BSCCO by $c$-axis
optical measurements at $T_c$ \cite{Kitano}.
The static out-of-plane dielectric function can be obtained from the sum rule
\be
\epsilon_c=\epsilon_1(0)+\frac{2}{\pi} \int_0^{\infty} \frac{\epsilon_2(\omega')}{\omega'}
{d\omega'},
\ee
where $\epsilon_2(\omega)$ is the dynamic component of $\epsilon_c(\omega)$ \cite{Pines}.

 Although, the $r$-dependence of the screened in-plane electrostatic interaction
between the condensed charges $q_i$ and $q_j$ is not perfectly $1/r$, we
approximate it with the expression $V_{eff}^{ab} \approx e^2/4 \pi \epsilon_0 
\epsilon_{ab} r$ as well, where screening is taken into account implicitly
via $\epsilon_{ab}$.
The lowest eigenvalue of $H$ given in Eq.~(\ref{hamilt}) is $E=\langle \Psi \vert H \vert \Psi \rangle$, where $\Psi(r_1,r_2)=\sum_{\vec{k}}^{'} g(k) e^{i\vec{k}\cdot\vec{r}_1} e^{i\vec{k}\cdot\vec{r}_2}$
being
the wavefunction of the interacting pair and $g(k)$ is a pair-correlation function
\cite{Tinkham}.
Then the energy of the lowest eigenstate of the condensate in the SC state is approximated by
using a pure dielectric form for the potential energy
\beqn
E_{tot}=T_{kin}^{ab}+E_c^{ab}+E_c^{IL}=\frac{\hbar^2}{2 m^{*}}\Delta \Psi(r_1,r_2) ~~~~~~~~~~~&& \\ \nonumber~~~~~~ +\frac{e^2}{4 \pi\epsilon_0} \biggm(\frac{1}{\epsilon_{ab}} \sum_{ij}
\frac{q_i^{(1)} q_j^{(1)}}{r_{ij}^{(1)}} 
~~~~~~~~&& \\ \nonumber
  +\frac{1}{\epsilon_c} \sum_{n=1}^{2} \sum_{m=2}^{\infty} \sum_{ij}^{2 N^2} \frac{q_i^{(n)} q_j^{(m)}}{r_{ij}^{(n,m)}}\biggm),~~~~
\eeqn
where $\hbar$ is the Planck constant, $m^{*} \simeq 4 m_e$ is the effective mass for the holes induced in the half-filled bands \cite{Stasio},
$n,m$ are layer indices,
$r_{ij}^{(1)}$ and $r_{ij}^{(n,m)}$ are the intra-layer and inter-layer point charge distances, respectively.
The most important components are the interactions with $r_{ij}^{(1,2)}$ (bilayer components),
however one has to sum up for the IL interactions with terms $r_{ij}^{(n,m)}$,
where $m=[2,\infty]$. Note that only the interactions of various layers
with the basel bilayer (FIG. 2) are considered along the $c$-axis in both directions (up and down).
The in-plane electrostatic screening $\epsilon_{ab}$ is completely separated from
the out-of-plane dielectric screening ($\epsilon_c$).
$q_i, q_j$ are the partial point charges/atoms in the $N \times N$ superlattice model at optimal doping.
\be
q_{i,j}=\pm \frac{4e}{3 N^2}, 
\label{partialcharges}
\ee
%where factor $2$ is due to the fact that
The factor $3$ is included in the denominator because each $CuO_2$ site consists of $3$ atoms
($q_{i,j} \approx \pm 0.053e$ at optimal charge separation, $N=5$).
For the sake of simplicity it is assumed that the charges are equally distributed
among Cu and O atoms within a $CuO_2$ site.
The number of the lattice sites in the characteristic superlattice
is $N^2$ and 
\be
\xi_{ab} \approx (N-1)a_0
\label{xiab}
\ee
where $a_0 \approx 3.88 \AA$ is the $ab$ lattice constant.
Note that the coherence length $\xi_{ab}$ can be given in $a_0$ unit and the real space period
$N$ can be compared with $\xi_{ab}$ via Eq.~(\ref{xiab}) (see also Fig. (1)).
$N_h=2e$ is the charge of the electron pair.
The anti-hole charges $q_i^{ahole}$ must satisfy the {\em charge sume rule} within a characteristic bilayer over a coherence area $\sim \xi_{ab}^2$
\be
\sum_{i=1}^{N^2} q_i^{ahole} = 4e,
\ee
where $q_i^{ahole}$ represents the anti-hole point charges.
Naturally the charge neutrality $\sum_{ij}^{2 N^2} (q_i^{hole}+q_j^{ahole})=0$ is also required. 
In this paper we study the lattice size $N \times N=5$, which we found
nearly optimal for a variety of cuprates. However, it can be interesting
to study the variation of the "characteristic" lattice size in different
cuprates as a function of various parameters (doping, pressure etc.). 
The kinetic energy of the boson condensate arises from
the hopping of anti-holes (hole-anti-hole exchange) between the adjacent sites within the sheets,
against the electrostatic background of the rest of the hole-anti-hole system. Single hole-anti-hole
exchange is forbidden since extraordinary repulsion occurs which leads to the redistribution
of the entire hole-anti-hole charge pattern. The collective intersite anti-hole hopping results in the
kinetic energy of the condensate.

\section{The condensation energy in the charge ordered bilayer superlattice model}

 Important consequence of the model outlined in this paper is that the condensation
energy ($U_0$) of the SC state can be calculated.
We will show here that
within our model
the primary source of $U_0$ is the net energy gain in IL Coulomb energy occurs due to  
the asymmetrical distribution of the condensed hole-charge
in the adjacent layers (see Fig. 2) below $T_c$.

 Per definition $U_0$ is the free energy difference of the normal state
and the superconducting state \cite{PWA}.
Therefore, taking the energy difference $U_0 \approx \Delta E_{tot}= \vert E_{tot}^{NS}-E_{tot}^{SC} \vert$,
using Eq. (7) we get
\be
U_0 \approx \Delta E^{ab} + \Delta E_c^{IL}.
\ee
Tha $ab$-plane contribution to the SC condensation energy $\Delta E^{ab}$ is expected to be
negligible at optimal doping.
That is because our Coulomb energy calculations indicate that the $ab$-plane Coulomb energy in the NS and in the SC state is nearly equal
and therefore $\Delta E_c^{ab} \approx 0$
and also the kinetic energy of the pair-condensate $E_{kin}^{ab}$ is much smaller
by several order of magnitude then IL coupling.
For instance we get for the prototypical $5 \times 5$ lattice $\sim -1.2 \times 10^{-20}$ J $ab$-Coulomb
energy both for the NS and for the SC COS. 
The COS of the NS is the one given in Fig~(\ref{fig_ns}).
The calculated $ab$-kinetic energy of the $5 \times 5$ codensate is $\sim 1.1 \times 10^{-29}$ J 
if we use the very simple formula ("electron in a box")
\be
T_{kin}^{ab} \approx \frac{\hbar^2}{2m^*} \frac{2e}{\xi_{ab}^2 \xi_c},
\ee
where $\xi_c$ is the $c$-axis coherence length.
The inter-layer Coulomb interaction (attraction) $E_c^{IL} \approx -1.3 \times 10^{-21}$ J for the $5 \times 5$ bilayer. 
It must be emphasized, however, that away from optimal doping
the Coulomb energy of the condensate might affect the magnitude of the condensation energy.
The abrupt jump of the measured condensation energy $U_0$ seen in the slightly overdoped regime
\cite{Tallon99} can be attributed to the increased $\Delta E^{ab}$ contribution to $U_0$. 

  In this article we restrict our study, however, to optimal doping and the study of the doping dependence of
$U_0$ is planned in the near future.
The main contribution to $U_0$ is then the IL energy gain $\Delta E_c^{IL}$ 
in the SC state can be given as follows,
\be
  U_0 \approx \vert E_c^{IL,NS}-E_c^{IL,SC} \vert.
\label{gain}
\ee
We can further simplify Eq.~(\ref{gain}) if
the NS contribution to Eq.~(\ref{gain}) is
$E_c^{IL,NS} \approx 0$, which holds if IL coupling is screened effectively in the NS (large density of the hole
content in the IL space, large $\epsilon_c(NS)$).
Again, if we assume $\epsilon_c \approx 33$, we get $E_c^{IL,NS} \approx +2 \times 10^{-18}$ J ($12.3$ eV)
for a charge ordered $5 \times 5$ lattice
which would be an extraordinarily large value for the IL Coulomb repulsion.
Assuming, however, $\epsilon_c \approx 10000$, we get the more realistic Coulomb repulsion of $E_c^{IL,NS} \approx +10^{-22}$ J.
Indeed there are measurements for cuprates which indicate large $\epsilon_c$ in the NS
in the range of $10^3$ to $10^5$ \cite{Cao}. 
Extraordinarily large dielectric constant has also been found recently
in perovskite materials \cite{Homes}.
Anyhow the "relaxation" of the huge IL Coulomb repulsion in hole-doped cuprates in the NS can not easily be
understood without the consideration of a large $\epsilon_c$.
Nevertheless experimental measurements and theoretical speculations suggest that
$\epsilon_c$ in cuprates and in other materials with perovskite structure is strongly temperature and
doping dependent \cite{Kitano}. 
The sharp decrease of $\epsilon_c$ below $T_c$ should also be explained within our bilayer
model by the pair-condensation of the hole-content to the sheets.
This is what can be seen in the $c$-axis optical spectra of cuprates \cite{Basov,Molegraaf}.
The lack of the $c$-axis plasma edge in the NS is an obvious experimental evidence of the
strong temperature dependence of $\epsilon_c$.
  Then we have
\be
2 (n+1) N^2 U_0 \approx E_c^{IL,SC},
\label{gain_sc}
\ee
where $E_c^{IL,SC}$ is the Coulomb energy gain in the SC state.
$U_0$ is the experimental condensation energy given per unit cell. 
%$E_{cond}$ is the condensation energy of the bilayer system with $2 N^2$ unit cells.
Eq.~(\ref{gain_sc}) is generalized for multilayer cuprates introducing
$n$, the number of $CuO_2$ layers within a multilayer block.

The hole-conductivity in the SC state is strictly 2D phenomenon, no
direct IL hopping of quasiparticles is considered within this model.
$T_c$ is mainly determined by the inter-plane distance
and by the static $c$-axis dielectric constant (the c-axis component of the dielectric tensor).
{\em An interesting feature of the $N \times N$ bilayer model is then that it is capable of retaining
the 2D character of superconductivity while $T_c$ is enhanced by 3D Coulomb interactions.}

\section{The condensation energy and $T_c$}

 There are number of evidences are available which suggest that HTSC occurs beyond
the BCS limit. In those materials which contain nearly isolated
single layers, such as $Bi_2Ba_2CuO_{6+\delta}$  ($T_c=20$ K \cite{Konstantin}) or 
superlattice structures (periodic artificially layered materials) such as O-doped
$(BaCuO_2)_2/(CaCuO_2)_n$ thin films, when $n=1$  \cite{Balestrino} and in
$YBa_2Cu_3O_{7-\delta}/PrBa_2Cu_3O_{7-\delta}$ (YBCO/PBCO) \cite{Plakida,Li,Terashima} the critical temperature is 
limited to $T_c \le 30 K$ (below the BCS limit), therefore these materials
are not considered as high-$T_c$ superconductors in this article.
In these compounds the IL distance is so large that the $CuO_2$ planes are
decoupled and the superconducting properties can be understood within the BCS
formalism. 
However, the multilayer Bi-compounds (with the same charge reservoir as in the single layer Bi2201),
the $(BaCuO_2)_2/(CaCuO_2)_n$ thin films, when $n \ge 2$ \cite{Balestrino}
or YBCO with 
thin $PrBa_2Cu_3O_{7-\delta}$ layer \cite{Plakida,Li,Terashima}  
exhibit HTSC.
Therefore it is worth to explain the enhancement of $T_c$ beyond the BCS limit
assuming other mechanism than the electron-phonon coupling.
Inter-layer Coulomb coupling can be a natural source of the condensation
energy and of HTSC.
Checkerboard-like charge pattern (COS) seen experimentally \cite{Hoffman,Lake,Howald} directly leads to IL energy gain and
to potential energy driven condensation energy if the hole-anti hole
charge pattern is asymmetrycally condensed to the adjacent layers (Fig 2).
Assuming that thermical equilibrium occurs at $T_c$ for the competing
charge ordered phases of the NS and the SC state the following equation
for the condensation energy can be formulated using Eq.~(\ref{gain_sc}),
\be
2 (n+1) N^2  U_0 \approx 2 (n+1) \biggm[ \frac{\xi_{ab}}{a_0}+1 \biggm]^2 U_0 \approx k_B T_c.
\label{kbtc}
\ee
At a first look this formula seems to be unusual because of the dependence
of the condensation energy on $T_c$. The available measurements of the
condensation energy on various cuprates show no correlation of $U_0$
with the critical temperature.

%%%%%%%%%%%%%%%%%%%%%
%%% TAB 1
%%%%%%%%%%%%%%%%%%%%
\begin{table}
%\center
\caption[]
{The calculated coherence length of the pair condensate
using the experimental condensation energies
of various cuprates and Eq.~(\ref{N}) at optimal doping.
}
{\scriptsize
\begin{tabular}{cccccc}
 & $T_c$ (K) & $k_B T_c$ (meV) & $U_0$ ($\mu eV/u.c.$)  & $\xi_{ab}^{calc} (a_0)$ & $\xi_{ab}^{exp} (a_0)$ \\ 
\hline
 LSCO   & 39  & 2.5 & $21^a$ &  $\sim 7$ &  $5-8^h$  
 \\
 Tl2201 & 85 & 7 & $100 \pm 20^b$ & $\sim 5$ & \\
 Hg1201 & 95 & 7.8 & $80-107^c$ &  $\sim 5$ & $5^c$ \\
  YBCO  & 92 & 7.5 & $110^d$ &  $\sim 3$ & $ 3-4^g $ \\
  Bi2212 & 89 & 7.3 & $107^e$ &  $\sim 4$ & $4-5^f$\\
%---------------------------------------------------------------
\end{tabular}}
{\small
$U_0$ is the measured condensation energy of various cuprates
in $\mu$ eV per unit cell at optimal doping.
$^a$ $U_0 \approx 2$ J/mol from \cite{Loram,Momono},
$^b$ \cite{Tsvetkov},
$^c$ $U_0 \approx 12-16$ mJ/g from \cite{Billon,Hgcond} and $\xi_{ab}$ from \cite{Thompson},
$^d$ $U_0 \approx 11$ J/mol from \cite{PWA2,Tallon99,TallonLoram},
$^e$ $U_0 \approx 10$ J/mol from \cite{Tallon02},
%$^g$ from \cite{coher}
$^f$ from recent measurements of Wang {\em et al.} \cite{Wang_sci},
$^g$ from \cite{Tinkham,Welp},
%$^g$ from \cite{Hoffman,Bi_coher},
$^h$ from \cite{Tinkham},
$\xi_{ab}^{calc}$ is calculated according to Eq.~(\ref{N}) and is also given in Table ~\ref{tab1} and $\xi_{ab}^{exp}$ is
the measured in-plane coherence length given in $a_0 \approx 3.9 \AA$.
The notations are as follows for the compounds:
LSCO ($La_{1.85}Sr_{0.15}CuO_4$),  
Tl2201 ($Tl_2Ba_2CuO_6$), 
Hg1201 ($HgBa_2CuO_{4+\delta}$),  
 YBCO ($YBa_2CuO_7$) and
 Bi2212 is $Bi_2Sr_2CaCu_2O_{8+\delta}$.
}
\label{tab1}
\end{table}

\begin{figure}[hbtp]
%\begin{figure}[!t]
\begin{center}
\includegraphics*[height=4.5cm,width=6.5cm]{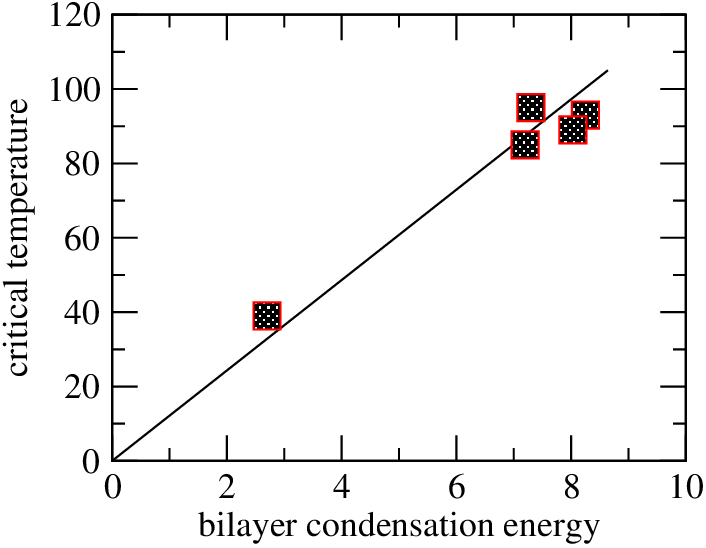}
\caption[]{
The critical temperature (K) at optimal doping as a function of the bilayer condensation energy
($2 (n+1) N^2 U_0$, meV) using Eq.~(\ref{kbtc}).
The real space period $N$ is directly related to $\xi_{ab}$ via Eq.~(\ref{xiab}).
The straight line is a linear fit to the data. The slope of the linear fit
is $k_B$ which is a strong evidence of Eq.~(\ref{kbtc}).
}
\label{cond_tc}
\end{center}
\end{figure}

One of the important goals of this paper, however, to show that correlation can indeed
be found with $T_c$ if $T_c$ is plotted against $2 N^2 U_0$ (the bilayer condensation energy).
 In other words {\em the condensation energy of the coherence bilayer-hole
system shows correlation with $T_c$}.
 In order to test the validity of Eq.~(\ref{kbtc}) we estimate the 
coherence length of the pair condensate using Eqs.~(\ref{kbtc})
and ~(\ref{xiab}) using only experimental data,
\be
\xi_{ab} \approx a_0 \biggm[ \sqrt{\frac{k_B T_c}{2 (n+1) U_0}}-1 \biggm]
\label{N}
\ee
The results are given in Table~\ref{tab1} as $\xi_{ab}^{calc}$ and compared with the available
measured $\xi_{ab}^{exp}$. The agreement is excellent which strongly suggests that Eq.~(\ref{N})
should also work for other cuprates.
 The validity of Eq.~(\ref{kbtc}) is clearly clarified in FIG~\ref{cond_tc} using mostly experimental
information for $\xi_{ab}$, $T_c$ and $U_0$ (values are given in Table I.).
For Tl2201 no measured $\xi_{ab}$ is found in the literature, therefore the estimated 
$\xi_{ab}$ is used (given in Table I.). In Section \ref{sec:diel} we will show that 
using this estimated value of $N=6$ we predict $\epsilon_c$ in nice agreement with
optical measurements for Tl2201.
{\em Remarkable feature of FIG~\ref{cond_tc} is that the slope of the linear fit to the
measured data points
precisely gives us $k_B$ which nicely confirms Eq.~(\ref{kbtc}).
}
The underlying physics of HTSC seems to be reflected by Eq.~(\ref{kbtc}):
the equation couples the observable quantities $T_c, U_0$ and $\xi_{ab}$.
We predict for the multilayer Hg-cuprates, Hg1212 ($T_c=126$ K) and Hg1223 ($T_c=135$ K) the condensation energies
$146$ and $117$ $\mu$eV/u.c. ($\xi_{ab} \approx 4 a_0$ \cite{Thompson}), respectively
using Eq.~(\ref{kbtc}).

%------------------------------------------------------

%------------------------------------------------------

\begin{figure}[hbtp]
%\begin{figure}[!t]
\begin{center}
\includegraphics*[height=4.5cm,width=6.5cm]{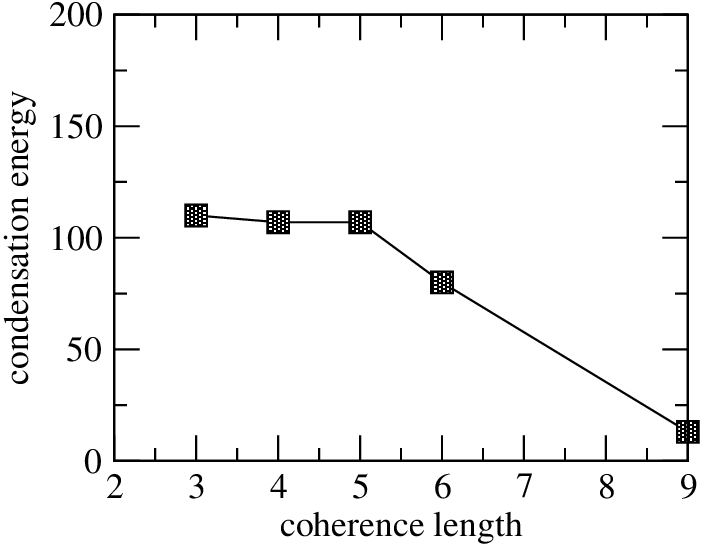}
\caption[]{
The measured condensation energy (given in Table I., $\mu$eV/u.c.) as a function of the calculated
coherence
 length ($a_0$) using Eq.~(\ref{N}).
}
\label{xi_cond}
\end{center}
\end{figure}

  Furthermore, we find correlation between the condensation energy $U_0$ and $N$: the larger $U_0$
is connected with smaller N (see also FIG~\ref{xi_cond}). {\em The stronger localization of the
pair-condensate wave function seems to lead to larger IL condensation energy and hence
to larger $T_c$.}
This is again an unexpected result, since the stronger localization of the Cooper
pairs should increase the Coulomb self-repulsion of the condensate and hence
should suppress $T_c$.
Note, however, that within a liquid-crystal-like COS the self-repulsion problem
is not crucial. The hole-anti-hole Coulomb interactions are attractive, although
cancelled by the intra-hole and intra-anti-hole repulsions (intra $CuO_2$ site repulsion,
).

  Interestingly the most localized coherence area ($\sim 4 a_0$) is provided by YBCO which is the
less anisotropic material among HTSC cuprates (the resisitivity ratio $\rho_{ab}/\rho_c \approx 200$
\cite{Plakida}).
The more anistropic Hg1201 gives us somewhat weaker localization of the condensate
wave-pocket for similar $T_c$. The comparison between various materials , however,
is much more complicated. In general we can say that $T_c$ is a function of the following
parameters considered in this study: $N$, $p$, the IL distance $d$ and $\epsilon_c$.
In section \ref{sec:diel} we further analyse this complex behaviour of $T_c$ focusing
on the calculated IL dielectric constant $\epsilon_c$.
The estimated $xi_{ab}$ values given in Tables~\ref{tab1} and ~\ref{tab2} of the bilayer COS are in close agreement with 
the experimental coherence areas $\xi_{ab}$ which supports the validity of
our basic Eq.~(\ref{kbtc}).

%%%%%%%%%%%%%%%%%%%%%
%%% TAB 2
%%%%%%%%%%%%%%%%%%%%
\begin{table}
%\center
\caption[]
{The calculated dielectric constant $\epsilon_c$ using Eq.~(\ref{epsc4})
in various cuprates as a function of the coherence length $\xi_{ab}$ of the charge ordered state.
}
{\scriptsize
\begin{tabular}{lccccccc}
 & $d (\AA)$ & $T_c (K)$ & $\xi_{ab}^{calc} (a_0)$ & $\epsilon_c$ & $\epsilon_c^{exp}$  \\ \hline
CaCuO2$^b$ &       4.64 &  110 & 4 & 80.8  &    \\ 
CaCuO2     &       3.19 &  89 &  7 & 83.5 & 
    \\ 
 LSCO &      6.65 &  39 & 5 &  27.9 & $23 \pm 3, 13.5^c$  \\
      &       &  & 6 & 27.3 &   \\
      &       &  & {\bf 7} & {\bf 11.3} &   \\
 Hg1201    &     9.5   & 95  & 5 & 33.5 & 34$^d$   \\
 Hg- (10 GPa) & $\sim 8.5$   & $\sim 105.$ & 4 & 27.6 & \\ 
 Hg- (20 GPa) & $\sim 8.2$   & $\sim 120.$ & 4 & 25.0 &   \\
 Tl2201 &    11.6 &  85 &  3 & 8.7 & 11.3$^e$ \\
        &         &     &  4 & 26.8  &  \\
        &         &     &  {\bf 5} & {\bf 13.0}  &  \\
        &         &     &   6 & 8.0  &  \\
 Bi2201 & 12.2 & 20  & 3 & 32.6 & $\sim 40^g$ \\ 
  &  &   & 4 & 110.8 &  \\ 
  &  &   & 5 & 10.1 & \\  
  YBCO  &   $8.5^h$  & 93  & {\bf 3} & {\bf 19.4} & $34^f,23.6^j$ \\ 
        &            &      & 4 & 31.1 &  \\ 
%---------------------------------------------------------------
\end{tabular}}
{\small
%where $N$ is the real space period of the characteristic square lattice
where $\xi_{ab}^{calc}$ is the estimated in-plane coherence length given in $a_0 \approx 3.9 \AA$.
The estimated $\xi_{ab}^{calc}$ is that value which is associated with that calculated $\epsilon_c$ 
which is in good agreemant with the experimental $\epsilon_c$.
the bold faced values are those which account
the best for comparison with experiment and are in accordance with the results
of Table ~\ref{tab1}.
$d$ is the $CuO_2$ plane to plane inter-layer distance in $\AA$ \cite{Mills}, $T_c$ is the experimental critical temperature.
$\epsilon_c$ is from Eq.~(\ref{epsc4}).
%$^a$ field effect doped material \cite{Schon}.
$^b$ $Ca_{0.3}Sr_{0.7}CuO2$ \cite{Azuma}, 
$^c$ \cite{epsilonc}, or from reflectivity measurements 
using Eq.~(\ref{plasma}), 
$\omega_p \approx 55 cm^{-1}$ \cite{Sarma}, $\lambda_c \approx 3 \mu$m \cite{Hgcond},
$^d$ experimental $\epsilon_c^{exp}$ values are taken from {\em Am Inst. of Phys. Handbook}, McGraw-Hill, Ed. D, E. Gray (1982),
the $\epsilon_c$ value of the ionic-background (Hg1201: BaO),
The pressure dependent $T_c$ values and IL distance date are taken from \cite{Novikov,Gao},
$^e$ from \cite{Tsvetkov}, 
$^g$ from \cite{Kitano},
$^h$ for YBCO the reduced IL $d=8.5 \AA$ is used instead of the
$c$-axis lattice constant (inter-bilayer block distance),
$^f$ from \cite{Ariosa},
%$^i$ from \cite{Bozovic},
$^j$ from reflectivity measurements: $\omega_p \approx 60 cm^{-1}$ \cite{Sarma}, $\lambda_c \approx 0.9 \mu$m \cite{Kitano2},
$\epsilon_c$ can be deduced using Eq.~(\ref{plasma}). 
%The notations are as follows for the compounds:
%LSCO ($La_{1.85}Sr_{0.15}CuO_4$), Hg1201 ($HgBa_2CuO_{4+\delta}$),  
% Bi2201 ($Bi_2Sr_2CuO_6$)
%  BSLCO ($Bi_2Sr
%_{1-x}La_xCuO_6$),
% NCCO ($Nd_{2-x}Ce_xCuO_4$) and YBCO ($YBa_2CuO_7$).
%and YBCO ($YBa_2CuO_{7-\delta}$).
}
\label{tab2}
\end{table}

%------------------------------------------------------

%------------------------------------------------------

\section{The critical temperature and the $c$-axis dielectric constant}
\label{sec:diel}

 The expression Eq.~(\ref{kbtc}) leads to the very simple formula for the critical temperature
\be
T_c (N,dq,d,\epsilon_c) \approx \frac{e^2}{4 \pi \epsilon_0 \epsilon_c k_B} \sum_{n=1}^{2}\sum_{m=2}^{N_l} \sum_{ij}^{2 N^2} \frac{q_i^{(n)} q_j^{(m)}}{r_{ij}^{(n,m)}}
\label{kbtc2}
\ee
where $N_l$ is the number of layers along the $c$-axis. When $N_l \rightarrow \infty$, bulk $T_c$ is calculated.
$T_c$ can also be calculated for thin films when $N_l$ is finite
and $\epsilon_c$ can also be derived
\be
\epsilon_c \approx \frac{e^2}{4 \pi \epsilon_0 k_B T_c} \sum_{n=1}^{2}\sum_{m=2}^{N_l} \sum_{ij}^{2 N^2} \frac{q_i^{(n)} q_j^{(m)}}{r_{ij}^{(n,m)}}
\label{epsc4}
\ee
where a $c$-axis average of $\epsilon_c$ is computed when $N_l \rightarrow \infty$.

 The calculation of the $c$-axis dielectric constants $\epsilon_c$ might provide
further evidences for Eq.~(\ref{kbtc}) when compared with the measured values \cite{Kitano,epsilonc,Ariosa}.
  In Table ~\ref{tab2} we have calculated the static dielectric function $\epsilon_c$ using Eq.~(\ref{epsc4})
and compared with the experimental impedance measurements \cite{Takayanagi}.
$\epsilon_c$ can also be extracted from the $c$-axis optical measurements using the
relation \cite{Tsvetkov,PWA2}
\be
\sqrt{\epsilon_c} = \frac{c}{\omega_p \lambda_c}, 
\label{plasma}
\ee
where $c, \omega_p$ and $\lambda_c$ are the speed of light, $c$-axis plasma frequency
and the $c$-axis penetration depth. 
The general conclusion can be drawn on the basis of the data depicted in Table~\ref{tab2}
that the $\epsilon_c$ values obtained from $c$-axis optical measurements are somewhat smaller
then the values obtained from impedance measurements \cite{Takayanagi}.
The reason for this is not clear.
Our calculations support the optical measurements {\em vs.} the impedance experiments
if $\epsilon_c$ is calculated at the experimental coherence length. 

 For the prototypical cuprate LSCO we get the value of $\epsilon_c=27.3$ which is
comparable with the experimental value of $23$ \cite{epsilonc} using $N=7$
which corresponds to $6 a_0 \approx 23.4 \AA$ ($\xi_{ab} \approx 20-30 \AA$ \cite{coher}).
The overall good agreement of the calculated $\epsilon_c$ with the measurements
is due to our finding that the IL charging energy is surprisingly in the
range of $k_B T_c$ when the real space period $N$ (coherence length) is reasonably chosen.

%\section{Relations to the superconducting gap and pairing}

  Not useless to note again the correlation in Table I. between $\xi_{ab}$ and $U_0$.
There seems to be a correlation between the real-space localization of the Cooper wave-function
and the SC energy gain $U_0$.

%------------------------------------------------------

\begin{figure}[hbtp]
%\begin{figure}[!t]
\begin{center}
\includegraphics*[height=4.5cm,width=6.5cm]{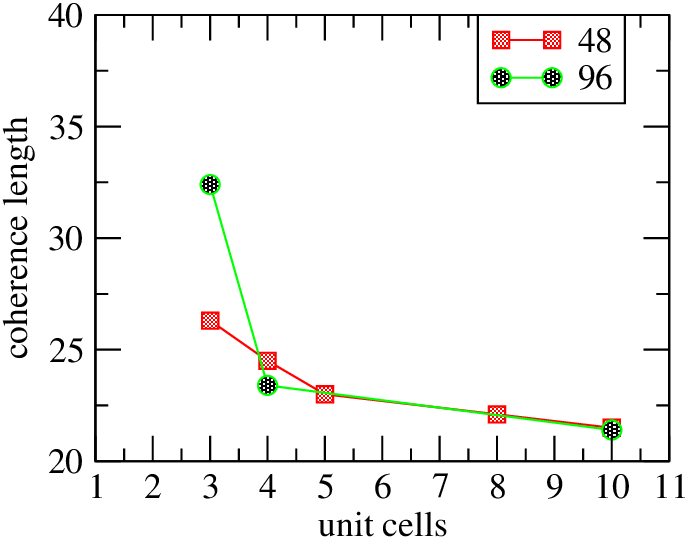}
\caption[]{
The coherence length $\xi_{ab}$ ($\AA$) as a function of the YBCO thickness
(in unit cells) at fixed thickness of the barrier (insulating) layer PBCO ($\sim 48$ and $96 \AA$)
reproduced from Table II. of ref. \cite{Yang}.
}
\label{fig_ybco_xi}
\end{center}
\end{figure}

%------------------------------------------------------

The real-space "shrinking" of the Cooper wave-function
must directly be related to the enhancement of pairing.
Within our model the primary source of increased pairing is the IL energy gain
(Coulomb induced pairing).

  Optimally doped YBCO/PBCO with decoupled layers shows no HTSC ($T_c \approx 20$ K) indicating
that optimal carrier concentration within isolated layers leads to probably BCS superconductivity
with low critical temperature.
Setting in IL coupling (decreasing the thickness of the PBCO phase) $T_c$ is enhanced due to the
increase of IL coupling \cite{Plakida,Li}.
Further experimental studies on artificially layered materials, such as
the measurement of $\xi_{ab}$ as a function of the thickness of the insulating phase PBCO should explain the importance of IL coupling
in cuprates. The considerable increase in $\xi_{ab}$ as a function of IL decoupling (increase in PBCO thickness or decrease in YBCO thickness) would be a strong
evidence for the model presented in sections III.-V.
Looking for such an article in the literature an interesting publication is found
which reveals our expectation \cite{Yang}. The extensive measurements of the upper critical field in $(YBCO/PBCO)_n$ superlattices
{\em indeed results in the increase of $\xi_{ab}$ as a function of decreased thickness of YBCO layers}.
$\xi_{ab}$ varies from $4-5 a_0$ up to $\sim 8 a_0$ when the thickness of the YBCO phase is
decreased from $10$ unit cells to $3$. Unfortunatelly this article contains no measurements
for $1-2$ unit cell thick samples. 
In FIG~\ref{fig_ybco_xi} we give the measured coherence length in YBCO/PBCO getting
the data from ref. \cite{Yang}.
We predict the rapid increase of $\xi_{ab}$ for dielectrically isolated single layers 
of YBCO for one or two-unit-cell thick samples which is the reminiscent of purely BCS
features.
{\em These striking results represent a strong evidence for the correlation between IL coupling
and the coherence length $\xi_{ab}$ which is predicted by the model presented here.
}
Strinkingly, the contraction of the Cooper wave-function (the strengthening of pairing) is strongly coupled to the
enhancement of IL Coulomb coupling.

%------------------------------------------------------

\begin{figure}[hbtp]
%\begin{figure}[!t]
\begin{center}
\includegraphics*[height=4.5cm,width=6.5cm]{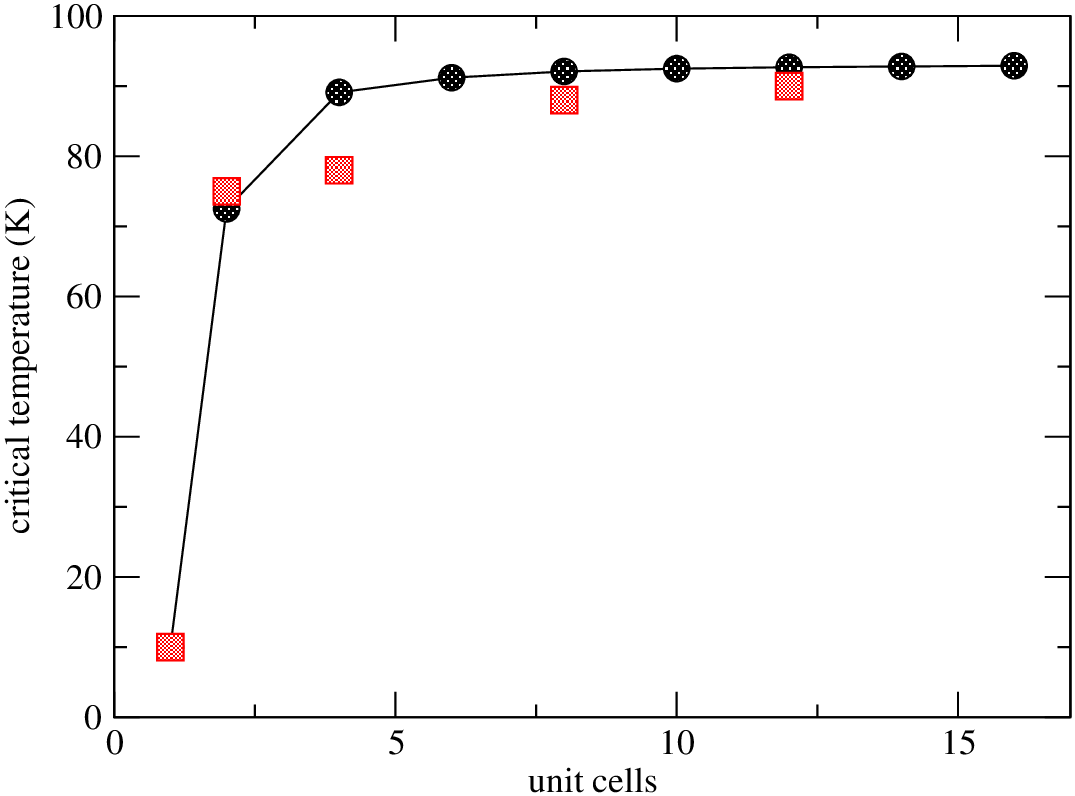}
\caption[]{The critical temperature $T_c$ (K, Eqs. ~(\ref{kbtc})-(\ref{epsc4})) {\em vs.}
the number of unit cells along the $c$-axis in YBCO using
the $4 \times 4$ model.  Circles and squares correspond to
the calculated and experimental values \cite{Li}.
}
\label{ybco_tc}
\end{center}
\end{figure}

%------------------------------------------------------

\section{Mutilayer and pressure effects}

 Multilayer and pressure effects on $T_c$ can also be discussed in terms of 
in-plane and out-of-plane localization of the hole charge.
%In the multilayer blocks $n_h/CuO_2$ is the half or third of the single layer value (in bilayer
%or trilayer systems). Therefore, there is no excess charge whithin the blocks in the
%interstitial regions, the confinement is robust, inter-layer charging energy peaks.
In multilayer systems $E_c^{IL}$ is composed of intra- and inter-block contributions.
We use the notation block for the multilayer parts $CaCu_2O_4$, $Ca_2Cu_3O_6$, etc. 
Eq.~(\ref{gain_sc}) can then be modified for multilayer cuprates as follows,
\be
 U_0 \approx E_c^{IL,intra} + E_c^{IL,inter},
\label{multil}
\ee
where
\be
E_c^{IL,intra}=(l-1) E_c^{IL}(d_{intra}, \epsilon_{intra}),
\ee
l denotes the number of layers.
Therefore, intra-block charging energy further enhances $T_c$ on top of the corresponding
single-layer inter-block value.
YBCO is a peculiar example of cuprates in which HTSC is purely coming from {\em inter-block
coupling} as it was explained on the basis of $YBa_2Cu_3O_{7-\delta}/PrBa_2Cu_3O_{7-\delta}$
superlattice structures \cite{Plakida}.
In YBCO the $Y(CuO_2)_2^{-2}$ bilayer alone does not show HTSC when isolated from each other
in artificially layered materials ($T_c \approx 20 K$).
The dependence of $T_c$ in YBCO on the number of unit cells along the $c$-axis
is calculated using Eq.~(\ref{kbtc2}) and the results are depicted on FIG~\ref{ybco_tc}.
The experimental points \cite{Li} are relatively well reproduced
indicating that $\sim 5$ unit cell thick thin film is already the reminiscent of
the bulk properties of cuprates.

 It is worth mentioning the {\em intercalation} experiment on Bi2212 \cite{Choy}.
Intercalation of I or organic molecules into the $(BiO)_2(SrO)_2$ layers of Bi2212 results
in no significant change in $T_c$.
Basically inter-layer intercalation is introduced to reduce interlayer coupling (test of inter-
layer theory, ILT) and to enhance anisotropy in SC properties.
Using film deposition technique, ultrathin films of Bi-Sr-Ca-Cu-O (Bi2212) have been
synthesized \cite{Saito}. The few-unit-cell-thick samples show HTSC similar to that of the bulk
material independently of the film thickness.
These results present great challenge to ILT \cite{PWA} and support
low-dimensional SC theories.

 In the light of the $N \times N$ model, however, it is not surprising that in Bi2212 the nearly isolated bilayers are 
are superconductors.
The single layer material Bi2201 ($Bi_2Sr_2CuO_6$) gives very low $T_c$ ($\approx 20 
K$) \cite{Konstantin},
which indicates that the dielectrics $(BiO)_2(SrO)_2$ strongly reduces IL coupling, indeed
the estimated
$\epsilon_c$ is quite large (Table ~\ref{tab2}) which is attributed to the weakly interacting $(BiO)_2$
bilayer (the BiO-BiO distance is $\sim 3.7 \AA$ \cite{Choy}). 
Dielectric constant measurements \cite{Varma} 
and the obtained small plasma frequency using $c$-axis optical measurements ($\omega \approx 5 cm^{-1}$)  
\cite{Sarma}
in Bi-compounds also provide relatively large
dielectric constants in accordance with our calculations.
The extremely large conduction anisotropy $\gamma \sim 10^6$ found in Bi-based cuprates also supports these findings \cite{Plakida}.
Nearest-neighbouring Cu-O planes in Bi-compounds are nearly insulated \cite{Ariosa}.
Therefore, in the multilayer Bi-compounds bilayer and trilayer blocks are responsible for
the high-$T_c$, inter-block coupling is negligible.
The multilayer Bi-compounds provide then an example for {\em pure
intra-block} HTSC. In these materials the multilayer-blocks are dielectrically isolated.
However, in the most of the cuprates $T_c$ is enhanced both via intra- and inter-block effects.
An important consequence of the bilayer $N \times N$ model is that {\em isolated single layers do not show HTSC},
coupling to next-nearest $CuO_2$ plane is essential in HTSC.
It is possible then to estimate $\epsilon_c$ for the bilayer block, since $E_c^{IL,inter} \approx 0$ in the Bi-compounds in  
Eq.~(\ref{multil}).
In the hypothetical bilayer system in Table ~\ref{tab2} ($Ca(CuO_2)_2, d=3.19 \AA, T_c=89 K$),
which is the building block of bilayer cuprates,
we find the value of $N=8$ accounting for realistic
$\epsilon_c$. $N=8$ is in accordance with the measured relatively
"large" coherence length of $\xi_{ab} \approx 27 \AA$ ($(N-1)a_0=7 a_0 \approx 27.3 \AA$) \cite{Bi_coher}.

 Also, upon pressure (p) the inter-layer spacing decreases, which increases $E_c^{IL}$ without
the increase of the hole content. Saturation of $T_c(p)$ is reached simply when 
increasing IL sterical repulsion starts to destabilize the system.  
In systems, such as YBCO or LSCO, negative or no pressure dependece of $T_c$ is found
\cite{Plakida} due to the short Cu-O apical distance ($d_{CuO} \approx 2.4 \AA$) which leads to already steric repulsion at ambient pressure and
to the weakening of HTSC. In these systems the net gain in charging energy is not enough 
to overcome steric repulsions at high pressures.
 We give the calculated $\epsilon_c$ values in Hg1201 for high pressures in Table I.
$\epsilon_c$ does not depend on the pressure and therefore the pressure dependence of $T_c$ can be given
by Eq. ~(\ref{kbtc2}).

 With these results, we are now in a position to reach the conclusion that the $N \times N$ model can readily account for at least certain physical properties
of multilayer cuprates, such as pressure,  doping and multilayer dependence of $T_c$.
Furthermore it is possible to make some estimations on the upper limit of $T_c$
using Eq.~(\ref{kbtc2}) for $T_c$. Assuming relatively small $\epsilon_c \approx 10$ and short inter-layer
spacing $d \approx 7.0 \AA$ we get the value of $T_c \approx 333 K$ for a 
strongly localized electron pair with $5 \times 5$ coherence area. Of course, we have no clear cut knowledge at this moment
on how the pair condensate wave-function spreads upon varying $\epsilon_c$ and $d$.
{\em Effective localization of the Cooper-pair wave-pocket could lead, however, to room temperature
superconductivity under proper structural and dielectric conditions if the model presented
above is applicable.}

 According to the stripe scenario, the charge ordered state of the coherence area can also be seen 
 in many cuprates as one-dimensional stripe order or charge density waves
 \cite{Howald}.
 The striped antiferromagnetic order found in LSCO by magnetic neutron 
 scattering experiments \cite{Lake} also implies a spin-density of periodicity $\sim 8 a$
 a reminiscent of the coherence area. 
 The incommensurate "checkerboard" patterns seen with a spatial periodicity of $\sim 8a_0$ in the vortex core of Bi2212 obtained by scanning tunneling microscopy \cite{Hoffman} is also consistent
 with our $N \times N$ hole-anti-hole charge ordered state where $N=8$ to $9$ in BSSCO. 

% Finally we mention the recent results of Bozovic {\em et al.} \cite{Bozovic}
%obtained for LSCO thin films under epitaxial strain. They reached the
%record $T_c=51.5 K$ for 15-unit-cell thick film of LSCO on $LaSrAlO_4$ substrate.
%The small variation of the $ab$- or $c$-axis lattice constants in our model accounts for only
%$1-2$ K increase in $T_c$.
%We explain the more then $10$ K enhancement of $T_c$ with the decrease of $\epsilon_c$
%(Table ~\ref{tab2}) under the conditions they used ($O_3$ annealing, epitaxial strain provided by the substrate).

%Moreover, the $\epsilon_c$ obtained for the field-effect doped $CaCuO_2$, is transferable
%to the chemically doped multilayer systems. 

%\newpage

\section{Conclusion}

 In this paper we studied the pair condensation and confinement of the hole-content on a $CuO_2$ superlattice layer
as a function of inter-layer distance and dielectric permittivity of the charge reservoir.  

\begin{itemize}
\item The assumption of a $2D \Leftrightarrow 3D$ quantum phase transition of the hole-content
at $T_c$
in HTSC materials is thought to be an important general feature of pair-condensation and is supported
by $c$-axis optical measurements and by first-principles calculations.
Our proposal is that the $c$-axis charge dynamics of the hole-content contributes
significantly to the condensation energy below $T_c$.
We find that
the inter-layer capacitance is temperature dependent in cuprates and therefore the
drop of the $c$-axis dielectric constant can be seen below $T_c$.
This is what leads to then the stabilization of the superconducting state {\em vs.}
the normal state.
\end{itemize}

\begin{itemize}
\item We have found that a pair condensate can be distributed on a $N \times N$ square lattice layer
in such a way that the lattice sites are filled by $q=\pm 0.16e$ condensed charge alternatively depending on the hole($+q$)-anti-hole($-q$) charge separation $dq$.
In this way a charge ordered state of the pair-condensate occurs 
with a "checkerboard" like pattern seen recently by experiment \cite{Hoffman}.
The phase separation of hole-electron pairs
(hole-anti-hole pairs) in this model is stabilized electrostatically.
The maximum charge separation is $dq \approx 0.32e$, if the optimal hole content $p_o \approx 0.16e$.
\end{itemize}

\begin{itemize}
\item In the adjacent layers the electron-hole pairs are distributed in such a way that complementer charge pattern occurs (inter-layer charge asymmetry) in order to maximize the inter-layer charging energy. 
A hole in one plane is neighbouring with an anti-hole in the other (see FIG 2.).
In this way we derived a charge ordered $N \times N$ correlated bilayer supercell model of the superconducting state
with inter-layer charge antisymmetry which directly leads to inter-layer Coulomb energy 
gain in the superconducting state.
The IL charge asymmetry could directly be tested experimentally using e.g. two-cell-thick
thin SC films, and the "checkerboard" STM images of both sides of the thin film could be
measured. A complementer "checkerboard" charge pattern would reveal our finding.
\end{itemize}

\begin{itemize}
\item The superlattice nature of the pair condensate is directly related to the
smallest size of the condensate wave-pocket which is remarkably comparable with
the measured in-plane coherence length of $\xi_{ab}=10-20 \AA$  ($4a_0 \approx 15.6 \AA$)
of single cuprates, where $a_0 \approx 3.9 \AA$ is the in-plane lattice constant.
The coherence area of the Cooper-pair wave-function in cuprates is strongly localized,
which is due to inter-layer charging effects.
\end{itemize}

\begin{itemize}
\item The bilayer model with $4e$ boson charge naturally implies the mass enhancement of $m^{*} \approx 4 m_e$
in accordance with measurements.
The calculated inter-layer charging energy is in the range of the experimental condensation
energy for the bilayer-hole system.
\end{itemize}

\begin{itemize}
\item
The static dielectric constant $\epsilon_c$ is calculated for a couple of cuprates
and compared with the available
experimental measurements. The general agreement is quite good indicating that
the pure inter-layer electrostatic model leads to proper description of
the static dielectric response of these layered materials. 
\end{itemize}

\begin{itemize}
\item The basic microscopic mechanism of HTSC is to be understood within the 
BCS-Eliashberg theory. The limiting critical temperature
for BCS-type superconductor is around $20 K$ as it was found for cuprates
with nearly isolated $CuO_2$ layers (BSCCO, YBCO/PBCO superlattices etc.).
The detailed study of this model showed that the inter-layer charging energy is proportional
to the thermal motion at $T_c$, if the $c$-axis dielectric constant $\epsilon_c$ and the coherence area is appropriately
chosen.
The correlation between the coherence area $\xi_{ab}$ and  inter-layer coupling is predicted
by our model. The stronger IL coupling leads to smaller $\xi_{ab}$.
This relation is evidenced by the measurements of $\xi_{ab}$ in YBCO/PBCO films
with varying YBCO thickness.
\end{itemize}

  If the physical picture derived from our model is correct, it should be a guide for 
further experimental studies aiming to improve SC in cuprates or in other materials.
This can be done by tuning the IL distance and $\epsilon_c$ (increasing the polarizability of the dielectric, hence decreasing $\epsilon_c$) in these materials.
The application of this model to other class of HTSC materials, such as fullerides or
$MgB_2$ is expected to be also effective.

\section{acknowledgement}

{\small
It is a privilige to thank M. Menyh\'ard for his continous support.
I greatly indebted to E. Sherman for reading the manuscript carefully
and for the helpful informations.
I would also like to thank for the helpful discussions with T. G. Kov\'acs.
Special thank should also be given for plenty of technical help
to A. Sulyok and to G. Tam\'as.
This work is supported by the OTKA grant F037710
from the Hungarian Academy of Sciences}
\\

%\newpage

\end{document}